\documentclass[11pt]{article}
\usepackage[margin=0.9in]{geometry}
\usepackage{amsmath}
\usepackage{amsfonts}
\usepackage{amssymb}
\usepackage{graphicx}
\usepackage{enumerate}
\usepackage{natbib}
\usepackage{url} 
\usepackage{multirow}
\usepackage{booktabs}
\usepackage{lipsum}
\usepackage[version=4]{mhchem} 
\usepackage{siunitx} 
\usepackage{bm} 
\usepackage{lineno} 
\usepackage{graphicx}
\usepackage{caption}
\usepackage{subcaption}
\usepackage{hyperref}
\usepackage{xcolor}
\usepackage{etoolbox}
\usepackage{changepage}

\newcommand\blfootnote[1]{%
  \begingroup
  \renewcommand\thefootnote{}\footnote{#1}%
  \addtocounter{footnote}{-1}%
  \endgroup
}

\newcommand{\indep}{\perp \!\!\! \perp}

\newcommand{\D}{\mathrm{d}}

\newtheorem{assumption}{Assumption}

\makeatletter
\def\blfootnote{\xdef\@thefnmark{}\@footnotetext}
\makeatother

\hypersetup{
    colorlinks=false,
    pdfborder={0 0 0},
    pdfborderstyle={/S/U/W 0},
}

\newcommand{\blind}{0}

\makeatletter
\patchcmd{\@maketitle}{\begin{center}}{\begin{adjustwidth}{-0.5in}{-0.5in}\begin{center}}{}{}
\patchcmd{\@maketitle}{\end{center}}{\end{center}\end{adjustwidth}}{}{}
\makeatother

\begin{document}

\bibliographystyle{agsm}


\def\spacingset#1{\renewcommand{\baselinestretch}
{#1}\small\normalsize} \spacingset{1} 


\if0\blind
{

  \title{\bf Causal health impacts of power plant emission controls under modeled and uncertain physical process interference} 
  \author{\small
	Nathan B. Wikle$^1$\thanks{Please direct correspondence to Nathan Wikle, Email: nathan-wikle@uiowa.edu} \; and \; Corwin M. Zigler$^2$
  }
\date{
    {\small
    $^1$Department of Statistics and Actuarial Science, University of Iowa \\
    $^2$Department of Statistics and Data Sciences, University of Texas at Austin\\[2ex]%
    }
    \today
}
  \maketitle
} \fi

\if1\blind
{

{
\title{\bf Causal health impacts of power plant emission controls under modeled and uncertain physical process interference} 
}

  \maketitle

} \fi

\spacingset{1} 


\begin{abstract}
Causal inference with spatial environmental data is often challenging due to the presence of interference: outcomes for observational units depend on some combination of local and non-local treatment. This is especially relevant when estimating the effect of power plant emissions controls on population health, as pollution exposure is dictated by (i) the location of point-source emissions, as well as (ii) the transport of pollutants across space via dynamic physical-chemical processes. In this work, we estimate the effectiveness of air quality interventions at coal-fired power plants in reducing two adverse health outcomes in Texas in 2016: pediatric asthma ED visits and Medicare all-cause mortality.  We develop methods for causal inference with interference when the underlying network structure is not known with certainty and instead must be estimated from ancillary data. Notably, uncertainty in the interference structure is propagated to the resulting causal effect estimates. We offer a Bayesian, spatial mechanistic model for the interference mapping which we combine with a flexible non-parametric outcome model to marginalize estimates of causal effects over uncertainty in the structure of interference. Our analysis finds some evidence that emissions controls at upwind power plants reduce asthma ED visits and all-cause mortality, however accounting for uncertainty in the interference renders the results largely inconclusive.
\end{abstract}

{\it Keywords:} causal inference, interference, air pollution, mechanistic models, BART


\spacingset{1.2}

\section{Introduction}
\label{sec:intro}

For decades, coal-fired power plant facilities have been a primary source of sulfur dioxide (\ce{SO2}) emissions in the United States \citep{Cullis1980, Orellano2021}. One of six ``criteria pollutants'' for which the US Environmental Protection Agency (EPA) sets national air quality standards \citep{EPA2013}, sulfur dioxide is notable for its harm to the environment and human health: \ce{SO2} emissions contribute to acidic deposition \citep{EPA2003} and are a key contributor to fine particulate matter (\ce{PM_{2.5}}), which has been associated with various adverse health outcomes, including respiratory and cardiovascular disease and death \citep{Pope2009}. Consequently, interventions which seek to limit \ce{SO2} emissions from power plants have been a regulatory priority in the United States for decades \citep{Dominici2014}, and recent studies have linked pollution derived from such emissions to impacts on all-cause mortality \citep{Henneman2023} and asthma \citep{Casey2020} among other health endpoints.

Flue-gas desulfurization (FGD) technologies, or \textit{scrubbers}, are one such intervention. FGD scrubbers are installed at coal-fired power plant facilities, where they remove (or ``scrub'') \ce{SO2} from the facility's combustion gases before they exit the smokestack. The reduction in \ce{SO2} after scrubber installation can be dramatic --- in some cases, desulfurization rates may exceed 95\% \citep{Li2022} --- which translates into substantial changes to ambient air quality in populations located downwind.  A careful understanding of the health benefits of scrubber installation is critical, both when evaluating their retrospective utility and when deciding which facilities should be targeted for future intervention.

In this paper, we seek to estimate the effect of scrubber installation on downwind health outcomes. Evaluating air quality interventions is challenging when treatment exposure is dictated, in part, by an underlying physical process \citep{ZiglerPapadogeorgou2021, Zigler2020}.  Pollutants are not stagnant, but rather are transported and deposited across space via physical processes such as wind and rain, a phenomenon known as \textit{air pollution transport}. Furthermore, \ce{SO2} emissions react with chemical constituents in the atmosphere to form particulate sulfate (\ce{SO4^{2-}}), contributing to the secondary formation of harmful ambient \ce{PM_{2.5}} \citep{Pope2009, Dominici2014}. Thus, the pathway connecting scrubber interventions to health outcomes is governed by the underlying transport and chemical reaction processes that produce harmful \ce{PM_{2.5}} from power plant emissions. 

A feature of this interconnectedness is the possible dependence of health outcomes at a given location on \textit{multiple} upwind treatments. In the causal inference literature, this is referred to as \textit{interference} \citep{Cox1958}, and its presence complicates effect estimation \citep{Sobel2006, Karwa2018}. Without additional assumptions on the structure of interference, the number of potential outcomes grows prohibitively large, rendering causal estimands meaningless \citep{Karwa2018}. One solution is to assume that the extent of interference is limited to some (known) network structure. Then, a unit's ``treatment'' can be deconstructed as two parts: a \textit{direct} intervention, which is assigned locally to the unit, and an \textit{indirect}, or \textit{neighborhood}, treatment exposure, which is defined via a function which maps the interventions of neighboring units to a scalar exposure level. This function has been called an \textit{exposure model} \citep{Karwa2018, vanderLaan2014}, \textit{exposure mapping} \citep{AronowSamii2017}, or \textit{interference mapping} \citep{ZiglerPapadogeorgou2021, Zigler2020}, and can be thought of as an extension of partial interference \citep{Sobel2006, HudgensHalloran2008, TchetgenThetgenVanderWeele2012, Forastiere2016} to more general interference structures. By restricting the treatment space to direct and indirect components, the number of potential outcomes is greatly reduced and causal estimands can be defined via contrasts of direct and indirect treatments \citep{AronowSamii2017, Karwa2018, Forastiere2021, Zigler2020}.

To date, almost all examples of causal inference with an exposure model have assumed that the network structure and form of the exposure model --- which together define the extent and strength of interference --- are known \textit{a priori}.  This may be justified when interference is assumed to result from social contacts, including contacts within neighborhoods \citep{HudgensHalloran2008, PerezHeydrich2014, LiuHudgens2014}, classrooms \citep{HongRaudenbush2006}, households \citep{BasseFeller2018}, and between buyers and sellers in an (online) marketplace \citep{Doudchenko2020}. In these settings, social networks are included in the data collection process and the exposure models are often convenient summaries (e.g., the assumption of stratified interference \citep{HudgensHalloran2008}). However, there is growing recognition that misspecified interference structures \citep{Savje2023} or heterogeneity in the type and strength of network interactions \citep{Qu2022} can impact the robustness of causal effect estimates, particularly in observational studies. Consequently, causal effect estimates may be sensitive to the amount of uncertainty surrounding the specified interference structure, and estimators which incorporate uncertainty in the interference structure may be desired.

In the context of air pollution, the network structure is not well defined by an obvious measure of contact or adjacency. Instead, the dependencies between outcome units and upwind treatments are dictated by the physical process itself \citep{Zigler2020}. Furthermore, air pollution transport is a \textit{stochastic} process, and uncertainty about this process implies uncertainty in the exposure model. Thus, the principal challenge for the statistician --- and the question this paper seeks to answer --- is whether knowledge and \textit{uncertainty} about this physical process can be incorporated into the definition of meaningful causal estimands and estimation procedures for observational studies with interference.

Whereas \cite{Zigler2020} approached a similar problem through the \textit{a priori} specification of a deterministic pollution transport model, we estimate the interference structure from available atmospheric pollutant and weather data.  We do so using a mechanistic, statistical model of atmospheric sulfate, originally developed by \cite{Wikle2022}, to characterize the dynamics of --- and uncertainty in --- the pollution transport process. Notably, the model’s mean and covariance structures are specified via an assumed advection-diffusion process, making it particularly well-suited to the problem of delineating how interventions at any given point might impact pollution at other locations. The mechanistic model is used to define a distribution of a (weighted) bipartite network linking scrubber interventions to outcome units, and the outcome’s exposure level is then characterized as a weighted average of the treatment status of upwind power plant facilities. Given its dependence on the learned spatio-temporal pollution transport dynamics, the exposure level represents the annual upwind treatment experienced at each outcome unit. Furthermore, our proposed estimation strategy simultaneously accounts for time-varying uncertainty in the transport process. Thus, the proposed interference network and exposure mapping provides an appropriate framework for the identification of the causal health impacts of emissions controls in populations susceptible to annual exposures to harmful particulates. Importantly, uncertainty in the estimated pollution transport process implies uncertainty in the specified interference structure. This paper proposes a two-stage Bayesian estimation procedure to propagate this uncertainty to the causal effect estimates; this can be viewed as relaxing the assumption of a (single) correctly specified interference structure that is commonly found in previous work on causal inference with network interference \citep{AronowSamii2017}.  This work joins \cite{Ohnishi2023} as one of the first examples of an inferred interference structure, in either the social network \citep{AronowSamii2017, Forastiere2021} or spatial interference \citep{Reich2021, ZiglerPapadogeorgou2021, Zigler2020, Wang2023} literature.  Notably, \cite{Ohnishi2023} use a Bayesian nonparametric method to estimate a latent exposure mapping within a set of experimental units, while our method uses a mechanistic model of treatment diffusion, fitted to auxiliary data, to estimate the interference structure in an observational study.

Alongside estimation of the structure of interference, we estimate causal effects of scrubbers on 1) pediatric asthma emergency department (ED) visits and 2) all-cause mortality among Medicare beneficiaries in Texas in 2016.  We adopt a flexible, Bayesian nonparametric model of the response surface; we use a log-linear Bayesian additive regression tree (BART) model for count data \citep{Murray2021}. This estimation procedure differs from the parametric procedures considered in \cite{Forastiere2021}, \cite{Zigler2020}, and \cite{Forastiere2018} --- we forgo propensity score modeling and instead rely on the nonparametric BART model for the purposes of confounding adjustment \citep{Hill2011, Hahn2020}. We advocate for a modular approach to Bayesian inference when estimating causal effects with a probabilistic exposure model --- this allows us to propagate the uncertainty in the interference structure to the causal effect estimates while avoiding model feedback \citep{Jacob2017}. We compare the performance of this estimator to one which ignores uncertainty in the interference structure --- unsurprisingly, incorporating interference uncertainty typically results in wider posterior credible intervals of the effect estimates. However, the total variance of the modular effect estimates can be decomposed into the sum of the variance due to the uncertain interference structure and the variance due to the outcome model. Thus, we can quantify how much of the uncertainty in the causal effect estimates is attributed to uncertainty in the interference structure. We find that modeling and acknowledging uncertainty in the interference structure has important implications for causal inferences about FGD scrubbers and both pediatric asthma ED visits and all-cause Medicare mortality. In particular, we find little evidence that FGD scrubbers affected the rate of asthma ED visits and all-cause Medicare mortality in Texas in 2016; this is especially true when the estimates include uncertainty in the interference structure.

\section{Scrubber Locations and Regional Health Outcomes}
\label{sec:data}
\begin{figure}[h]
\centering
     \begin{subfigure}[b]{0.48\textwidth}
         \centering
         \includegraphics[width=\textwidth]{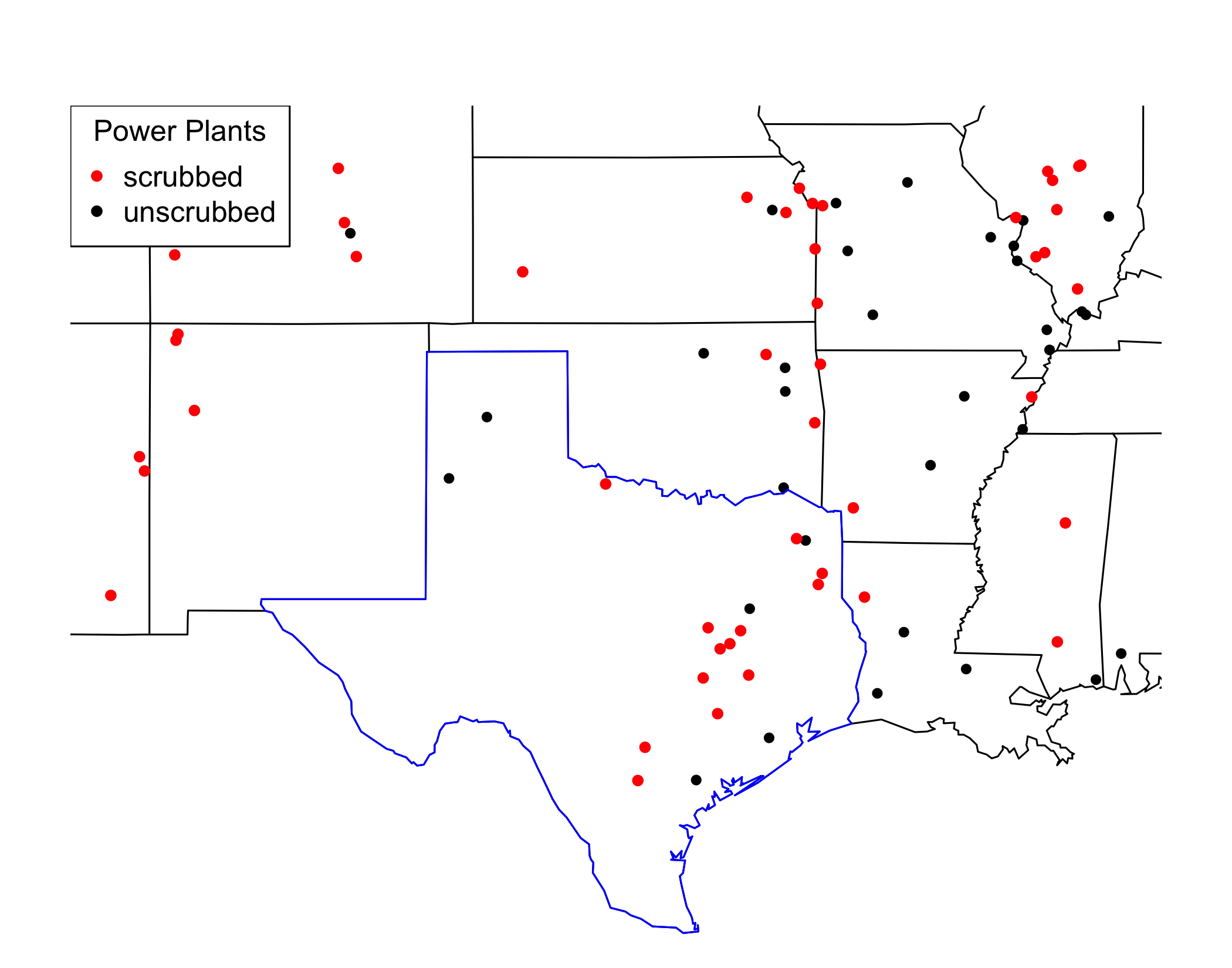}
         \caption{Power plants (scrubbers shown in red)}
         \label{fig:scrubbers}
     \end{subfigure}
     \hfill
     \begin{subfigure}[b]{0.48\textwidth}
         \centering
         \includegraphics[width=\textwidth]{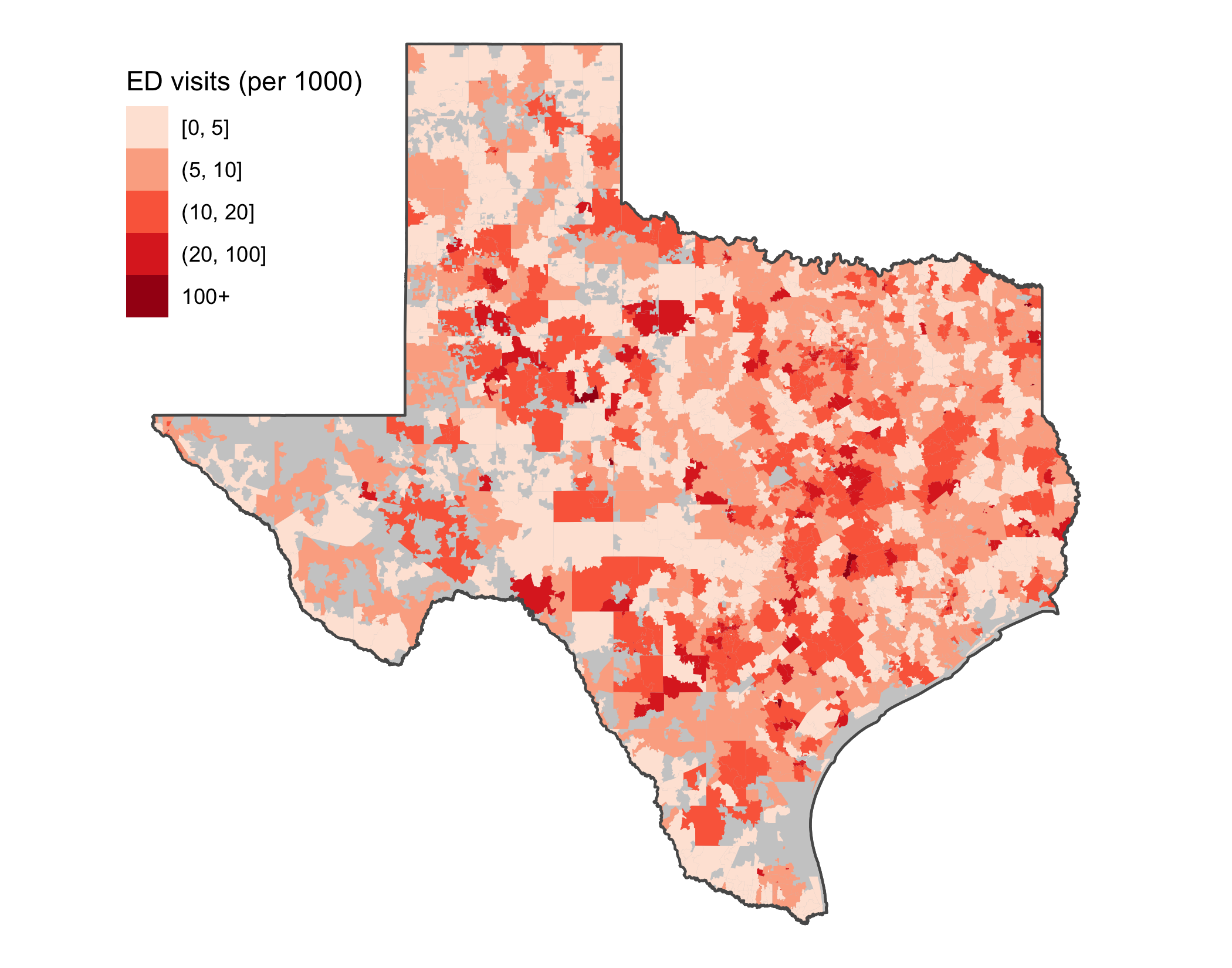}
         \caption{Asthma ED visits in 2016}
         \label{fig:asthma-rate}
     \end{subfigure}
\caption{Treatment and outcome data from our study of air quality interventions, including (a) the location and scrubber status of 81 coal-fired power plant facilities in the central US in 2016 (the interventional units), and (b) the rate of pediatric asthma ED visits in Texas in 2016, aggregated by ZCTA (the outcome units).}
\end{figure}

In 2016, there were 81 coal-fired power plant facilities operating in the area of the central United States expected to possibly influence air pollution exposure in Texas --- of those facilities, 48 were outfitted with scrubbers (Figure~\ref{fig:scrubbers}). Data on power plants were obtained from the US Environmental Protection Agency (EPA) Air Markets Program Database (AMPD) \citep{EPA2016}, and include several important facility-level variables, including annual \ce{SO2} emissions totals, operating time, and total heat input. 

We consider two health outcomes, both previously linked to exposure to \ce{PM_{2.5}} from power plants: pediatric asthma emergency department (ED) visits \citep{Garcia2021} and all-cause mortality among Medicare beneficiaries \citep{Dominici2014}. The asthma ED data were obtained from the Texas Health Care Information Collection (THCIC) Emergency Department Research Data File \citep{THCIC}, and include counts of pediatric asthma ED visits in Texas, aggregated into annual counts (or rates) according to patient ZIP code of residence. Note that ZIP codes were converted to their corresponding US Census ZIP Code Tabulation Areas (ZCTAs) using a data crosswalk provided by \cite{UDSMapper}. The Medicare data were obtained from the Center for Medicare and Medicaid Services, and were similarly aggregated to annual counts of all-cause mortality for each Texas ZCTA. Figure~\ref{fig:asthma-rate} shows the 2016 pediatric asthma rate for 1,935 Texas ZCTAs; we are unable to include a plot of the Medicare outcomes due to privacy constraints. For both outcomes, we restrict our analysis to ED visits/Medicare deaths that occurred in 2016, matching the temporal resolution of the power plant emissions and \ce{SO4^{2-}} data.

\section{Bipartite Causal Inference with Interference}
\label{sec:bipartiteCI}

\subsection{Potential Outcomes for Bipartite Causal Inference}

By themselves, the scrubber and outcome data in Section~\ref{sec:data} are of limited use --- there is not an obvious correspondence between treatment assignment and outcome unit; scrubbers are assigned to points in space (power plant facilities) while health outcomes are reported by ZCTA. In fact, if we assume the existence of interference, we are left wondering which upwind power plant facilities have the largest impact on air pollution exposure within a particular ZCTA. \cite{ZiglerPapadogeorgou2021} have styled this problem as \textit{bipartite causal inference with interference}, in reference to the directional bipartite network possibly linking the set of treatment locations to a separate set of outcome units. Note that this approach is distinct from efforts which seek to estimate the spillover effect as a function of the distance from a spatially-located treatment, such as those considered in \cite{Wang2023} and \cite{Pollmann2023}. Importantly, the bipartite framework naturally accommodates the inclusion of known physical processes which contribute to spillover effects; we briefly describe the framework as it applies to the analysis of air quality interventions.

Let $\mathcal{J} = \{1, \dots, J\}$ be the collection of power plant facilities in our study --- we refer to these as the \textit{interventional units}. The treatment status of interventional unit $j$ is denoted with $S_j \in \{0, 1\}$, where $1$ indicates the existence of an FGD scrubber at facility $j$, and $0$ otherwise. The treatment vector $\bm{S} = (S_1, \dots, S_J)$ represents the treatments assigned to all interventional units in $\mathcal{J}$. Finally, $\bm{s} \in \mathcal{S}(\mathcal{J})$ denotes a particular realization of the treatment vector $\bm{S}$, where $\mathcal{S}(\mathcal{J})$ is the space of all possible treatment vectors. 

Similarly, let $\mathcal{N} = \{1, \dots, N\}$ denote the set of \textit{outcome units} at which the health endpoints of interest are observed.  In our analysis $\mathcal{N}$ is the collection of 1,935 Texas ZCTAs at which asthma ED visits and Medicare all-cause mortality were reported in 2016. Let $Y_i$ denote the observed health outcome at ZCTA $i$ in 2016.  Together, the interventional units, $\mathcal{J}$, and outcome units, $\mathcal{N}$, form two disjoint sets of vertices of a bipartite graph. Each set of vertices have associated covariates, which we denote as $\bm{X}^{int}$ and $\bm{X}^{out}$ for the interventional and outcome units, respectively.

The challenge of bipartite causal inference, then, is in the definition of the edge set between $\mathcal{J}$ and $\mathcal{N}$. Notably, because of the difference in spatial support between the interventional points (power plant locations) and outcome units (ZCTAs), there is not an obvious one-to-one mapping between $\mathcal{J}$ and $\mathcal{N}$ \citep{ZiglerPapadogeorgou2021, Zigler2020}. Furthermore, the presence of interference implies the existence of \textit{multiple} edges between an outcome unit $i \in \mathcal{N}$ and the intervention set, $\mathcal{J}$. Thus, without additional structural assumptions on the extent of interference, unit $i$'s potential outcome, $Y_i(\bm{s})$ --- the outcome that would be observed at unit $i \in \mathcal{N}$ had the treatment vector $\bm{s} \in \mathcal{S}(\mathcal{J})$ been assigned --- depends on the treatment assignment at \textit{every} interventional point $j \in \mathcal{J}$. This means that the number of potential outcomes is very large: for binary interventions, $S_j \in \{0, 1\}$, there are $2^J$ potential outcomes (i.e., $|\{Y_i(\bm{s})\}_{\bm{s} \in \mathcal{S}(\mathcal{J})}| = 2^J$), only one of which is observed.

\subsection{Defining an Exposure Model}

The large number of potential outcomes can be alleviated if interference can be characterized or approximated by an exposure model, $g_i: \mathcal{S}(\mathcal{J}) \rightarrow \mathcal{G}$, which maps the treatment space $\mathcal{S}(\mathcal{J})$ to a set of scalar exposure values, $\mathcal{G}$. Then, relevant causal estimands are defined as contrasts of the exposure values $G_i \in \mathcal{G}$ (and possibly some other function of the treatment assignment, such as the treatment status of the nearest power plant). This approach has been used with some success when estimating spillover effects on social networks, where interference is assumed to occur due to social contacts between individuals \citep{AronowSamii2017, Forastiere2021}. In the social network setting, the exposure mapping is defined as a function of the network topology --- for example, \cite{Forastiere2021} define $g_i$ as the proportion of unit $i$'s friends who have received treatment. Note that the social network is typically assumed to be static and measured without error in the data collection process.

\cite{Zigler2020} consider a similar approach in the bipartite setting, wherein they define a weighted adjacency matrix, $T$, connecting interventional and outcome units. The adjacency matrix defines the edge weights of a bipartite graph, $\mathbb{G} = (\mathcal{J}, \mathcal{N}, T)$. The elements of the adjacency matrix, $T_{ij}$, can be interpreted as the relative influence of interventional unit $j$ on outcome unit $i$; larger values of $T_{ij}$ indicate particularly influential power plants connected to outcome unit $i$. 

The interference structure is then characterized by two components --- a \textit{direct} and an \textit{indirect} treatment. Let $j^{*}_{(i)}$ denote the interventional unit (power plant) that is geographically closest to outcome unit $i$. We call this the \textit{key-associated} unit \citep{Zigler2020}, and similarly define $Z_i = S_{j^{*}_{(i)}}$ to be the key-associated treatment for outcome unit $i$. Thus, $Z_i = 1$ if the nearest power plant to ZCTA $i$ is scrubbed, and 0 otherwise. This can be thought of as a ``direct'' treatment on outcome $i$, which is desirable for two reasons. First, it matches the convention in the social network literature \citep{Forastiere2021}, where every unit receives a corresponding direct treatment assignment. Second, and more importantly, it defines the scrubber status of the power plant which is often of most regulatory or community interest; causal estimands can then be defined which quantify the effect of scrubber installation at the nearest power plant.

The remaining power plant treatments, $\bm{S}_{-j^{*}_{(i)}}$, are then mapped to an ``indirect'', or \textit{upwind}, exposure level, $G_i$. This is accomplished with an exposure model, $g_i(\bm{S}_{-j^{*}_{(i)}}, T): \{0, 1\}^{J-1} \rightarrow \mathcal{G}_i$, where $G_i \in \mathcal{G}_i$ denotes a scalar upwind treatment value. Importantly, $g_i(\bm{S}_{-j^{*}_{(i)}}, T)$ is a function of the non-key-associated treatments, $\bm{S}_{-j^{*}_{(i)}}$, and the bipartite network's adjacency matrix, $T$. The utility of the exposure model is determined by how well the adjacency matrix, $T$, links the interventional units (i.e., power plant facilities) to the outcome units most affected by their treatment. If the structure of interference is well-specified by $T$, then $G_i$ provides an interpretable summary of the upwind treatment status for outcome unit $i$.

The potential outcomes notation can now be extended to bipartite settings with upwind interference. In particular, when combined with the familiar no multiple versions of treatment (consistency) assumption \citep{Forastiere2021}, the following upwind interference assumption serves as a modified version of the stable unit treatment value assumption (SUTVA) \citep{Zigler2020}:
\begin{assumption}[Upwind Interference] \label{asmp:upwind_interference}
For a fixed $T$ and exposure model $g_i(\cdot, T)$, $\forall i \in \mathcal{N}$, and for any two $(\bm{S}, \bm{S}') \in \mathcal{S}(\mathcal{J})$ such that $Z_i = Z_i'$ and $G_i = G_i'$, the following equality holds:
$$Y_i(\bm{S}) = Y_i(Z_i, G_i) = Y_i(Z_i', G_i') = Y_i(\bm{S}').$$
\end{assumption}
In other words, we have assumed that the interference structure is completely characterized by an outcome unit's direct and indirect treatment levels, $Z_i$ and $G_i$. In the power plant example, this implies that the health outcome of interest --- i.e., the rate of pediatric asthma ED visits or all-cause mortality among Medicare beneficiaries --- in ZCTA $i$ would be the same under any two scrubber allocations, $(\bm{S}, \bm{S}')$, so long as the key-associated and upwind exposure levels $(Z_i, G_i)$ remain the same.

\subsection{Causal Estimands for Bipartite Interference}

The upwind interference assumption is of critical importance --- meaningful causal estimands can now be defined as simple contrasts of $Y_i(Z_i, G_i)$. Two estimands are immediately relevant: a ``direct'' effect, which considers the effect of an intervention at the key-associated interventional unit, and an ``indirect'' or ``upwind'' effect, characterizing the spillover effect from treatments at all other interventional units, as summarized by the upwind exposure level, $G_i$. We formalize these estimands as follows.

Let $\mu(z,g)$ denote the marginal mean of the potential outcome, $Y_i(z,g)$:
\begin{equation}
    \mu(z,g) = E(Y_i(Z_i = z, G_i = g)).
\end{equation}
This is the expected value of the potential outcome of the $i$th outcome unit when the key-associated intervention has treatment value $z$ and the upwind treatment level is $g$. Thus, $\mu(z,g)$ is an average dose-response function (or surface) for different levels of $Z_i$ and $G_i$. For convenience, we have implicitly assumed that the potential outcome $Y_i(z,g)$ is defined for all values of $g \in \mathcal{G}$ and for all outcome units $i \in \mathcal{N}$; the definition can be suitably modified when the interference structure prohibits certain values of $g$ for some $i$ \citep{Forastiere2021}. Furthermore, we have construed the potential outcomes for each unit as random variables; in Section~\ref{sec:effects_est} we discuss how to estimate these unobserved random variables using (nonparametric) Bayesian models.

Using the above notation, we define the direct effect as
\begin{equation}
    DE(g) = \mu(1,g) - \mu(0,g), \label{eq:direct_effect}
\end{equation}
the expected effect of treating the key-associated unit, while holding the upwind units' treatment level constant. Notably, $DE(g)$ is a function of $g$, which allows for the possibility that the effect of a scrubber on the closest power plant is heterogeneous with respect to the scrubber intensity of upwind facilities. If desired, we can marginalize over $g$ to define an \textit{average direct effect} of the key-associated treatment, $ADE = \sum_{g \in \mathcal{G}} DE(g) \hat{P}(g)$, where $\hat{P}(g)$ denotes the estimated distribution of $G_i$ across the study domain.

Similarly, we define the indirect (or upwind) effect as 
\begin{equation}
    IE(z,g) = \mu(z,g) - \mu(z, g_{min}). \label{eq:indirect_effect}
\end{equation}
Here, $IE(g;z)$ denotes the expected change in outcome as the upwind treatment exposure changes from some baseline value, $g_{min}$, to an exposure level $g$, while the key-associated treatment level is fixed at $z$. Thus, $IE(z,g)$ represents the expected spillover effect of the upwind (i.e., non-key-associated) treatments, which may vary with $g$ and $z$. The choice of baseline value $g_{min}$ is application specific, for example, it could be set to zero, or to the minimum level of $G_i$ observed with the data. Finally, the \textit{average indirect effect} is defined as $AIE(z) = \sum_{g \in \mathcal{G}} IE(z,g) \hat{P}(g)$.

\section{Estimating the Interference Structure with a Mechanistic Spatial Model}
\label{sec:sulfate_model}

The utility of the bipartite potential outcomes framework, as outlined in Section~\ref{sec:bipartiteCI}, hinges on the specification of the network adjacency matrix ($T$) so that the exposure model ($g_i(\cdot, T)$) carries sufficient interpretability to define meaningful direct and indirect causal effects. Of course, the central challenge of this work is that $T$ cannot be observed --- there is no preordained network connecting power plant facilities to ZCTAs. Furthermore, simple proximity-based assignment of ZCTAs to power plants would grossly simplify the process of long-range pollution transport. Instead, our knowledge of the underlying mechanisms governing pollution transport --- and human exposure to harmful particulate matter --- can be leveraged to \textit{estimate} $T$. 

Estimation of the interference structure consists of two steps: first, we use a mechanistic statistical model of sulfate pollution to estimate the long-range pollution transport dynamics; we then extract $T$ from the fitted model. Notably, the estimated uncertainty about the process dynamics induces a probability distribution on $T$, which in turn leads to uncertainty in $G_i$. It's reasonable to ask what is gained from estimating (with uncertainty) the interference structure from observed pollution concentrations, rather than using output from a deterministic chemical transport model. For example, \cite{Zigler2020} characterize $T$ using output from a reduced-complexity atmospheric model (HyADS, see \cite{Henneman2019}), and there exists a rich body of work developing deterministic physical models of air pollution. However, these deterministic models can be computationally intensive, particularly at the spatial resolution considered in our analysis, and are themselves associated with uncertainty that cannot be easily quantified. Furthermore, pollution transport is an inherently random process --- small variations in factors such as wind velocity, precipitation, and chemical reactants contribute to the realized pollution exposure --- and it seems desirable that process uncertainty should propagate both to the interference structure and to the causal estimates.

\subsection{A mechanistic model of annual sulfate}
\label{sec:mech_model}

\begin{figure}[!ht]
\centering
     \begin{subfigure}[b]{0.48\textwidth}
         \centering
         \includegraphics[width=\textwidth]{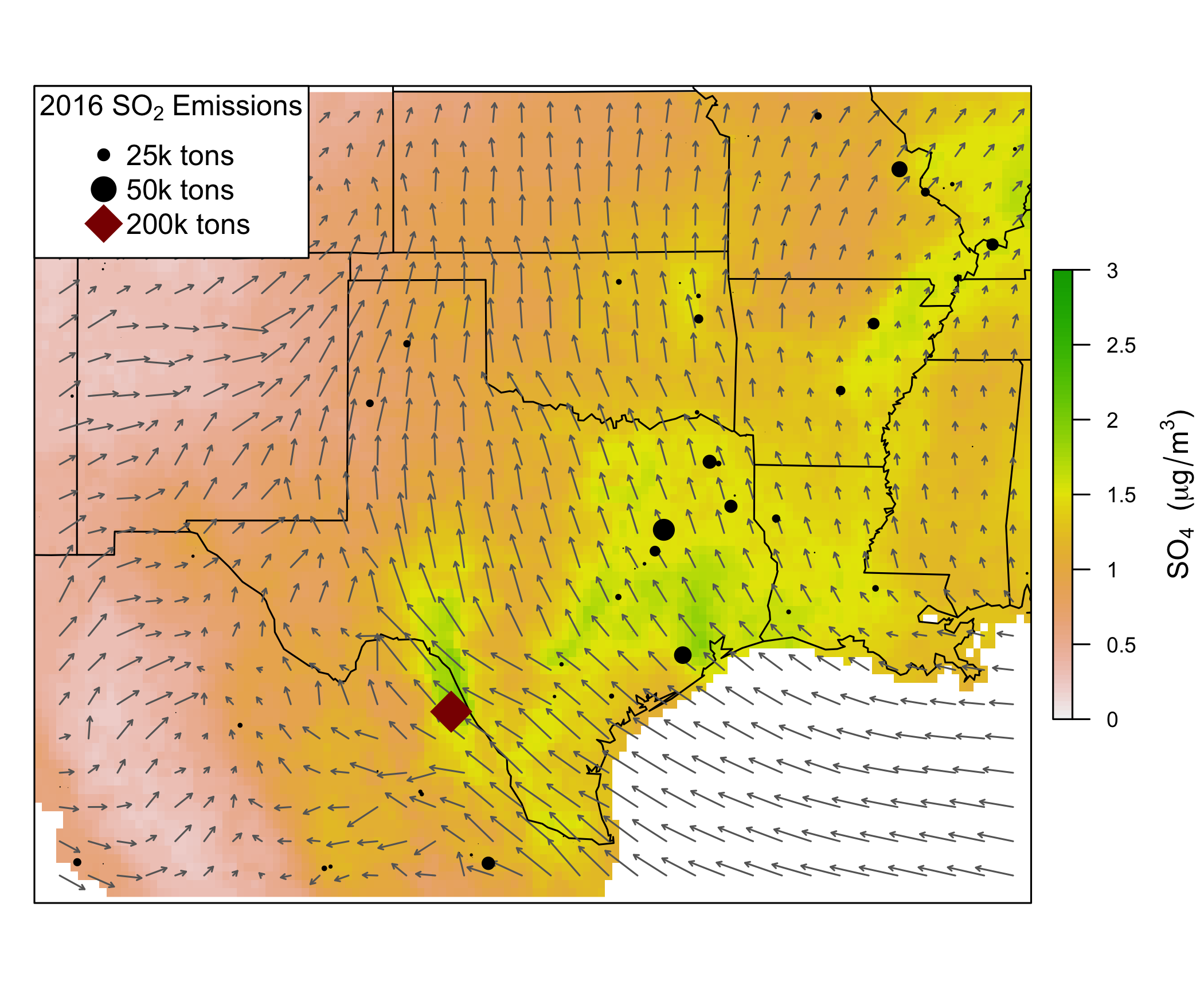}
         \caption{Average 2016 sulfate concentrations}
         \label{fig:obs_so4}
     \end{subfigure}
     \hfill
     \begin{subfigure}[b]{0.48\textwidth}
         \centering
         \includegraphics[width=\textwidth]{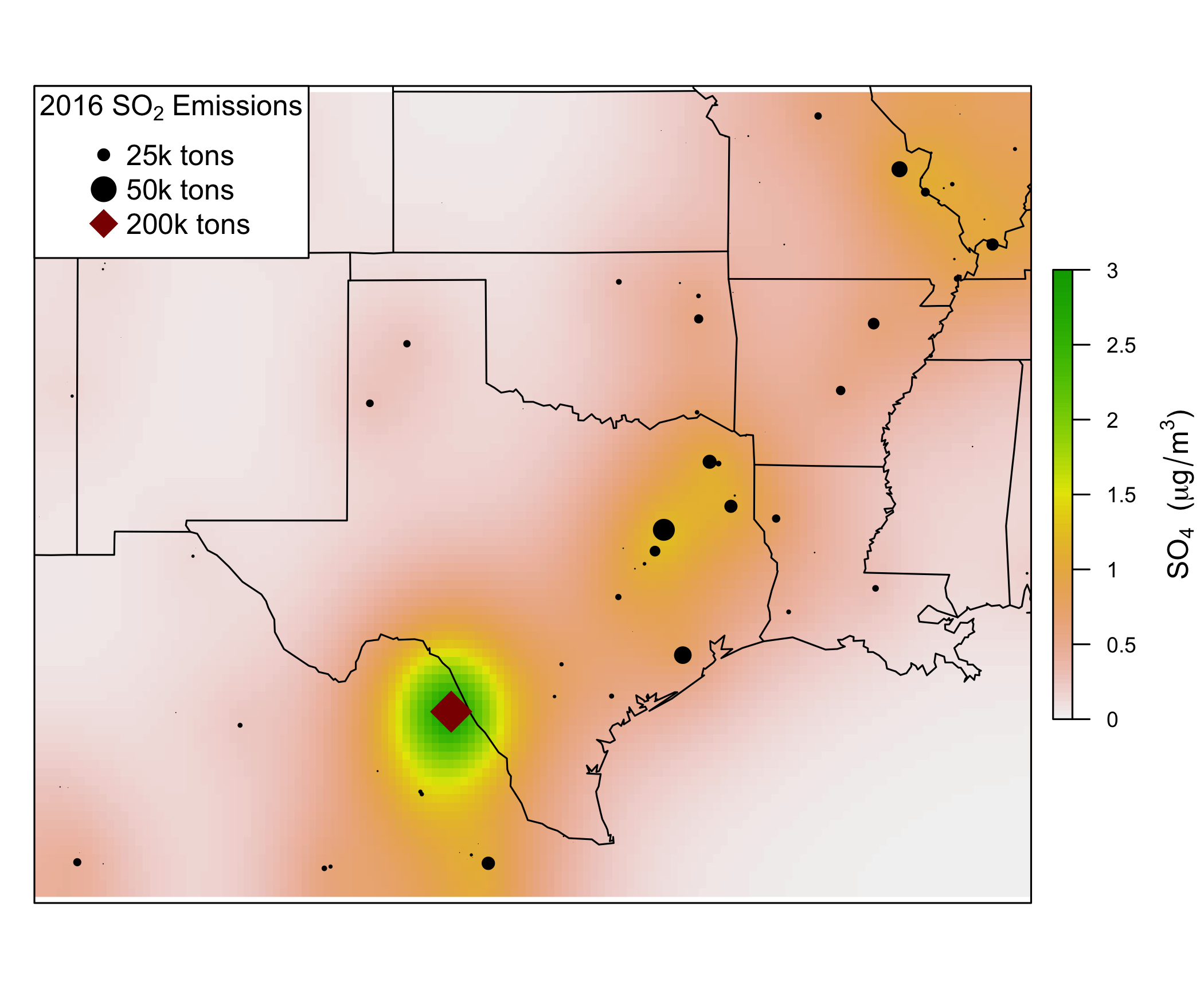}
         \caption{Estimated mean \ce{SO4} due to power plant emissions}
         \label{fig:est_so4}
     \end{subfigure}
\caption{A summary of the learned air pollution transport dynamics, including (a) the observed 2016 average sulfate concentrations, overlaid with power plant \ce{SO2} emissions totals and average 2016 wind velocities (at a height of 10 m above ground), and (b) the estimated 2016 sulfate concentrations attributed to coal-fired power plant emissions.} 
\label{fig:pollution_transport}
\end{figure}

We define a mechanistic model of annual sulfate concentrations in the US, in which the transport process is estimated, with uncertainty, from three 2016 data sources: coal-fired power plant emissions totals, average atmospheric sulfate concentrations, and average yearly wind velocity. The data are shown in Figure~\ref{fig:obs_so4}, and were obtained from the EPA AMPD \citep{EPA2016}, the Atmospheric Composition Analysis Group \citep{vanDonkelaar2019}, and the NCEP/NCAR reanalysis database \citep{Kalnay1996}, respectively. Most statistical approaches for modeling air pollution exposures are phenomenological, which would focus in this case on accurately interpolating a surface from observed sulfate measurements \citep{vanDonkelaar2019, Guan2020}, without characterizing how pollution moves from one location to another. Consequently, their ability to link power plant emissions to expected pollution concentrations is limited. In contrast, we use the class of mechanistic statistical models developed by \cite{Wikle2022}, in which known process dynamics of pollution transport, defined as a linear stochastic partial differential equation (SPDE), are approximated in discrete space as a multivariate Ornstein-Uhlenbeck (OU) process. This model can be easily fit to the data in Figure~\ref{fig:obs_so4}, and provides inference on how changes in \ce{SO2} emissions at a specific point dictate changes across the entire sulfate surface. We briefly describe the model, deferring to \cite{Wikle2022} for details.
  
Let $\eta(\bm{s}, t)$ denote the concentration of atmospheric \ce{SO4^{2-}} at location $\bm{s} \in \mathcal{D} \subset \mathbb{R}^2$ and time $t \in \mathbb{R}^{+}$, and let $\nu(\bm{s}, t)$ denote the corresponding local \ce{SO2} concentration. We model pollution transport as a coupled advection-diffusion process:
\begin{equation} \label{eq:so2_transport}
\D \nu(\bm{s}, t) = \big( -\mathcal{L}_{\bm{\theta}}(\bm{s}, t) \, \nu(\bm{s}, t) + R_{\bm{\theta}}(\bm{s}, t) \big) \D t
\end{equation}
\begin{equation} \label{eq:so4_transport}
\D \eta(\bm{s}, t) = \big( -\mathcal{A}_{\bm{\theta}}(\bm{s}, t) \, \eta(\bm{s}, t) + \theta_3 \, \nu(\bm{s}, t) \big) \D t + \xi(\bm{s}, t)
\end{equation} 
Equation~\eqref{eq:so2_transport} approximates how emissions from power plants move in space and time, defining the transport of \ce{SO2} across space: $-\mathcal{L}_{\bm{\theta}}(\bm{s},t) = (\Delta_{\theta_1} - \bm{w} \cdot \nabla_{\theta_2} - \theta_3)$ is an advection-diffusion operator, where $\Delta_{\theta_1}$ denotes homogeneous diffusion with rate $\theta_1$ ($\Delta$ is the Laplace operator in $\mathbb{R}^2$), $\bm{w} \cdot \nabla_{\theta_2}$ denotes advection due to wind ($\bm{w}$ is the wind velocity field, $\theta_2$ is a constant rate of advection), and $\theta_3$ is the rate at which \ce{SO2} is oxidized into \ce{SO4^{2-}}. Finally, $R_{\bm{\theta}}(\bm{s}, t)$ represents sources of \ce{SO2} emitted at location $\bm{s}$ and time $t$. Note that the operator $\mathcal{L}_{\bm{\theta}}(\bm{s}, t)$ acts on $\nu(\bm{s}, t)$, the local concentration of \ce{SO2}, while the emissions sources, $R_{\bm{\theta}}(\bm{s}, t)$, are independent of $\nu(\bm{s}, t)$.

Equation~\eqref{eq:so4_transport} approximates how ambient sulfate pollution moves in time and space, defining a similar advection-diffusion process for \ce{SO4^{2-}}, with three important distinctions. First, the advection-diffusion operator $-\mathcal{A}_{\bm{\theta}}(\bm{s},t) = (\Delta_{\theta_1} - \bm{w} \cdot \nabla_{\theta_2} - \delta)$ is almost identical to $\mathcal{L}_{\bm{\theta}}(\bm{s},t)$, however, $\theta_3$ (the oxidation rate of \ce{SO2} $\rightarrow$ \ce{SO4^{2-}}) has been replaced with $\delta$, the rate of atmospheric deposition of \ce{SO4^{2-}}. Second, the source term, $R_{\bm{\theta}}(\bm{s}, t)$, has been replaced with $\theta_3 \nu(\bm{s}, t)$, which accounts for the reaction of \ce{SO2} into \ce{SO4^{2-}}. Third, we have introduced a space-time Gaussian noise process, $\xi(\bm{s}, t)$, to account for space-time varying sources and sinks of \ce{SO4^{2-}} that were otherwise unspecified in the model; the addition of $\xi(\bm{s}, t)$ makes \eqref{eq:so4_transport} an SPDE. 

Together, \eqref{eq:so2_transport} and \eqref{eq:so4_transport} model a physical system in which (i) \ce{SO2} is emitted from the point locations of operating power plants, (ii) \ce{SO2} emissions are advected across space by wind, (iii) \ce{SO2} reacts into \ce{SO4^{2-}}, which is itself advected across space before eventual atmospheric deposition, and (iv) the system is inherently random, better reflecting possible fluctuations due to changes in weather, elevation, or chemical constituents that are otherwise unaccounted for in the model. However, solving the SPDE remains a challenge. A solution to this problem, as outlined by \cite{Wikle2022}, is to approximate \eqref{eq:so2_transport} and \eqref{eq:so4_transport} in discrete space with an Ornstein-Uhlenbeck (OU) process \citep{UhlenbeckOrnstein1930}. Then, the distributional properties of the OU process are leveraged to define a Gaussian likelihood model for spatial data, where the process dynamics, as specified in \eqref{eq:so2_transport} and \eqref{eq:so4_transport}, determine the mean and covariance structure of the model. 

For the sake of brevity, the discretization details have been confined to the Supplementary Material. Instead, we focus on the resulting mechanistic statistical model: let $\bar{\bm{\eta}}$ denote the observed annual average \ce{SO4^{2-}} concentrations, as shown in Figure~\ref{fig:obs_so4}. The resulting likelihood model is
\begin{equation} 
\bar{\bm{\eta}} \sim N\bigg( \beta_0 + \bm{\mu}_{\bm{\theta}}(\bm{R}), \, \bm{\Sigma}_{\bm{\theta}} \bigg), \label{eq:so4_likelihood}
\end{equation}
where $\bm{\mu}_{\bm{\theta}}(\bm{R})$ denotes the expected annual average sulfate (as specified by the advection-diffusion process in \eqref{eq:so2_transport} and \eqref{eq:so4_transport}) attributable to the annual coal-fired power plant \ce{SO2} emissions totals, $\bm{R}$, and $\bm{\Sigma}_{\bm{\theta}}$ denotes the covariance matrix, which has a simultaneous autoregressive (SAR) structure defined according to the specified dynamic process (see the attached Supplementary Material), as well as \cite{Wikle2022}, for detailed descriptions of $\bm{\mu}_{\bm{\theta}}$, $\bm{\Sigma}_{\bm{\theta}}$, and the choice of priors for $\bm{\theta}$). Note the dependence of $\bm{\mu}_{\bm{\theta}}$ and $\bm{\Sigma}_{\bm{\theta}}$ on $\bm{\theta}$, the (unknown) parameters from the advection-diffusion process. Finally, a small difference with \cite{Wikle2022} is the inclusion of $\beta_0$, which represents ``background'' \ce{SO4^{2-}} from emissions sources outside the study area.  

A likelihood derived from \eqref{eq:so4_likelihood} was fitted to the 2016 annual average sulfate data (Figure~\ref{fig:obs_so4}). We used a Bayesian approach for inference: independent half-normal and exponential priors were chosen for $\bm{\theta}$ and $\beta_0$, and posterior samples were obtained via Metropolis-Hastings MCMC. Figure~\ref{fig:est_so4} shows $\hat{\mu}_{\bar{\bm{\theta}}}$, the estimated posterior mean sulfate concentrations attributable to coal-fired power plant \ce{SO2} emissions in 2016, without background sources. Finally, we note that the estimated mean annual sulfate concentrations are comparable to estimates from a deterministic, reduced-complexity atmospheric model, with broadly similar spatial patterns exhibiting more spatial diffusion and higher estimated levels of \ce{SO4^{2-}} (Supplementary Material).  

Figure~\ref{fig:pollution_transport} illustrates two important advantages of this spatial model. First, because it is mechanistic, it can be used to estimate sulfate concentrations under counterfactual emissions scenarios by estimating how a change in emissions at any power plant impacts \ce{SO4^{2-}} across the entire region. This will prove useful when defining the network adjacency matrix, $T$, and distinguishes it from alternative phenomenological statistical models of air pollutants, which focus on spatial interpolation rather than mechanism \citep{vanDonkelaar2019, Guan2020}. Second, it includes estimates and uncertainty quantification about the process parameters $\bm{\theta}$. In contrast, more complex numerical models of air pollution, such as chemical transport models, plume models, and their reduced form hybrids \citep{Foley2014, Henneman2019} are often deterministic (and computationally expensive). Thus, this model's ability to infer (simplified) process dynamics and stochastic fluctuations from data makes it an ideal candidate for defining the dependence between regional outcome units (ZCTAs) and upwind power plant facilities.

\subsection{Estimating the interference structure with uncertainty}
\label{sec:exposure_model}

The learned dynamics in Section~\ref{sec:mech_model} can be used to define a probabilistic network adjacency matrix, $T$, connecting the interventional units, $\mathcal{J}$, to the outcome units, $\mathcal{N}$. For a given outcome unit $i$, we are interested in identifying the upwind power plants that have the greatest \textit{potential} influence on pollution exposure. Consequently, we would expect scrubbers placed at these influential power plants to have a larger effect on pollution exposure. We characterize the relative influence of an interventional unit $j$ on outcome unit $i$ through a source-receptor (SR) matrix, which we define using $\bm{\mu}_{\bm{\theta}}$, the mean function of our mechanistic spatial model. Let $D_i \subset \mathbb{R}^2$ denote ZCTA $i$'s geographic boundary and let $\bm{r}_j$ denote an emissions scenario in which power plant $j$ emits 1000 tons of \ce{SO2} in 2016, and all other coal-fired power plant \ce{SO2} emissions are set to zero. Then, $T_{ij}$ is defined as
\begin{equation}
T_{ij} = \frac{1}{|D_i|} \int_{D_i} \bm{\mu}_{\bm{\theta}}(\bm{r}_j),    \label{eq:sr_mat}
\end{equation}
the expected average \ce{SO4^{2-}} concentration in ZCTA $i$ per 1000 tons \ce{SO2} emitted from facility $j$. This calculation is repeated for all outcome unites, $i \in \mathcal{N}$, and for all interventional units, $j \in \mathcal{J}$. The resulting SR matrix defines the edge weights of a bipartite graph, $\mathbb{G} = (\mathcal{J}, \mathcal{N}, T)$; larger values of $T_{ij}$ indicate particularly influential power plants connected to outcome unit $i$. Characterizing an SR matrix by sequentially evaluating emissions from individual sources parallels traditional efforts used to evaluate power plant impacts, but typical reliance on chemical transport modeling for this purpose proves computationally prohibitive for a large number of individual source-population links \citep{Buonocore2014}. 

The definition of $T$ in \eqref{eq:sr_mat} is dependent on the specified process parameters, $\bm{\theta}$, and the uncertainty around these parameters can be propagated to $T$. In particular, if $\bm{\theta}^{(k)} \sim \pi(\bm{\theta} | \bar{\bm{\eta}}, \bm{R})$ is a sample from the posterior of $\bm{\theta}$, then $T^{(k)}$ denotes the associated adjacency matrix estimated with $\bm{\mu}_{\bm{\theta}^{(k)}}$. Repeating this for all MCMC samples of $\bm{\theta}$ provides samples from $\pi(T | \bar{\bm{\eta}}, R)$, the distribution over $T$. In short, we have estimated an interference structure with uncertainty: the pollution transport dynamics are learned from the available sulfate data, and the learned dynamics define the distribution of a weighted network connecting power plant facilities to ZCTAs.  

\subsection{Defining an exposure model}

Finally, we use $T$ to define a (probabilistic) exposure model, $g_i$. As discussed in Section~\ref{sec:bipartiteCI}, each outcome unit $i$ is assigned a direct treatment value, $Z_i = S_{j^{*}_{(i)}}$, which denotes the scrubber status of the key-associated interventional unit, $j^{*}_{(i)}$. The treatment vector of the remaining interventional units, $\bm{S}_{-j^{*}_{(i)}}$, is mapped to an indirect treatment level, $G_i$. Let 
\begin{equation}
G_i \equiv g_i(\bm{S}_{-j^{*}_{(i)}}, T) = \sum_{j \neq j^{*}_{(i)}} T_{ij} S_j  / T^{*}_{i\cdot},
\end{equation}
where $T^{*}_{i\cdot} = \sum_{j \neq j^{*}_{(i)}} T_{ij}$ is the (weighted) degree of node $i$. Note that $G_i$ represents the weighted proportion of treated upwind (i.e., non-key-associated) power plants; treated facilities that are more influential (as measured by $T$) contribute to larger values of $G_i$. Once again, the estimated uncertainty in $\bm{\theta}$ can be propagated to $G_i$ by repeatedly calculating $G_i^{(k)}$ using different samples from $\bm{\theta}^{(k)} \sim \pi(\bm{\theta} | \bar{\bm{\eta}}, \bm{R})$. Direct and indirect causal estimands can now be defined as contrasts of potential outcomes, $Y_i(Z_i, G_i)$, as shown in Section~\ref{sec:bipartiteCI}. 

\section{Estimating Causal Effects using Poisson Regression with BART}
\label{sec:effects_est}

In the simplified case where $G_i$ is fixed and known, identification of the direct and indirect treatment effects, $DE(g)$ and $IE(z,g)$, requires additional assumptions of unconfoundedness. In particular, the presence of two treatment components to characterize interference necessitates an extension of the familiar ignorable treatment assumption \citep{Zigler2020}:
\begin{assumption}[Ignorability of Joint Treatment Assignment] \label{Ignorability of Joint Treatment Assignment} \label{asmp:ignorability}
$$Y_i(z,g) \indep Z_i, G_i \,\, | \,\, \bm{X}^{out}_i, \, \bm{h}(\{\bm{X}^{int}_{j}\}_{j \in \mathcal{J}}), \,\, \forall z \in \{0, 1\}, \forall g \in \mathcal{G}_i, \forall i \in \mathcal{N}$$
\end{assumption}
In other words, the assignment of the key-associated treatment, $Z_i$, and the upwind treatment, $G_i$, are independent of the potential outcomes, conditional on a set of local (outcome unit) covariates, $\bm{X}^{out}_{i}$, and a (possibly multivariate) function of covariates associated with the interventional units, $\bm{h}(\{\bm{X}^{int}_j\}_{j \in \mathcal{J}})$. The need to condition on both outcome and interventional unit covariates is a distinct feature of bipartite networks, as discussed in \cite{Zigler2020}, and the exact specification of $\bm{X}^{out}_{i}$ and $\bm{h}(\{\bm{X}^{int}_j\}_{j \in \mathcal{J}})$ should be guided by application-specific knowledge of potential confounders. In addition, identification of $DE(g)$ and $IE(z,g)$ requires a common assumption of overlap,
$$0 < p\left(Z_i = z, G_i = g \, | \, \bm{X}_i \right) < 1,$$
where $\bm{X}_i \equiv \{\bm{X}^{out}_{i}, \bm{h}(\{\bm{X}^{int}_j\}_{j \in \mathcal{J}}) \}$. Given these two assumptions, 
\begin{equation}
    \mu(z,g) = \frac{1}{n} \sum_{i=1}^{n} E( Y_i | X_i = x_i, Z_i = z, G_i = g),
\end{equation} 
and the estimation of $\mu(z,g)$ has been simplified to a more familiar task: estimating the response surface, 
\begin{equation}
    E(Y_i \, | \, Z_i = z, G_i = g, \bm{X}_{i} = \bm{x}). \label{eq:resp_surface}
\end{equation}

In particular, we estimate \eqref{eq:resp_surface} with a log-linear Bayesian additive regression trees (BART) model for count data. First introduced by \cite{Murray2021}, log-linear BART is an extension of BART for continuous data with Gaussian errors \citep{Chipman2010} to regression models with Gamma-Poisson likelihoods. Letting $Y_i$ denote the observed count outcome data, the log-linear BART model for Poisson regression is defined as
\begin{equation}
    Y_i \, | \, \bm{x}_i, z_i, g_i \sim \text{Pois}\left(\mu_{0i} f_{\bm{\alpha}}(\bm{x}_i, z_i, g_i)\right), \label{eq:pois_lik}
\end{equation}
\begin{equation}
    \log f_{\bm{\alpha}} = \sum_{k = 1}^{m} h(\bm{x}_i, z_i, g_i ; \alpha_h). \label{eq:log_BART}
\end{equation}
Here, $\mu_{0i} f_{\bm{\alpha}}(\bm{x}_i, z_i, g_i)$ is the conditional expected value of $Y_i$, $\mu_{0i}$ denotes a fixed offset (such as ZCTA population size), and $\log f_{\bm{\alpha}}$ is a flexible sum of $m$ independent regression trees, $h(\bm{x}_i, z_i, g_i ; \alpha_h)$, where $\alpha_h$ denotes the tree parameterization \citep{Murray2021}. Fitting this model is non-trivial;  \cite{Murray2021} proposes a mixture of generalized inverse Gaussian distributions as a conjugate prior for the leaf parameters, which allows for a block MCMC update of the tree structure and leaf parameters. The BART-based model --- appropriately extended to count data --- retains the flexible nonparametric structures that have led to the increased popularity of BART for causal inference \citep{Hill2011, Dorie2019, Hahn2020}, and also represents a departure from previous methods for interference based on propensity score modeling \citep{Forastiere2018, Forastiere2021, Zigler2020}. In particular, BART's documented strengths include the ability to prioritize among a potentially high-dimensional set of covariates and nonlinarities among them.

\section{Propagating Interference Uncertainty to the Causal Estimates}
\label{sec:prop_unc}

Finally, we consider how uncertainty in the estimated interference structure might be propagated to the causal effect estimates. This premise assumes that the assumptions of Section~\ref{sec:effects_est} hold, at least approximately, for every value of $G$ simulated from the posterior of the process model. As discussed in Section~\ref{sec:exposure_model}, the sulfate model --- and the corresponding interference structure --- was based on our underlying knowledge of the physical sulfate transport process. Consequently, we want to avoid unwanted ``feedback'' from the health outcome model influencing inference for the treatment model parameters. Furthermore, misspecification of either model may lead feedback from one model to ``corrupt'' the other, introducing bias or misleading uncertainty quantification of the causal effects. 

To combat the possibility of unwanted feedback, we advocate for a \textit{modular} approach to inference, in which the interference structure (i.e., the sulfate model) is estimated without inclusion of the health outcomes data. As discussed in \cite{Jacob2017}, Bayesian modularization schemes are increasingly relevant in many applied settings, including studies with uncertain air pollution estimates \citep{Blangiardo2016, Comess2022}. Rather than targeting the full Bayesian posterior, we instead obtain inference using the ``cut function,''
\begin{equation}
    \pi^{*}(\bm{\alpha}, \bm{\theta} \; | \; \bm{y}, \bar{\bm{\eta}}) = \pi(\bm{\alpha} \; | \; \bm{y}, \bm{\theta}) \pi(\bm{\theta} | \bar{\bm{\eta}}). \label{eq:cut_function} 
\end{equation}
Note that \eqref{eq:cut_function} is not equivalent to the full Bayesian posterior, as the dependence of $\bm{\theta}$ on $\bm{y}$ has been ``cut.'' As discussed in \cite{Plummer2014}, sampling from \eqref{eq:cut_function} is non-trivial. Typically, inference proceeds via a computationally-intensive, multiple imputation approach: $K$ samples of $\bm{\theta}$ are first obtained via MCMC from the first module, $\pi(\bm{\theta} \, | \,  \bar{\bm{\eta}})$. Then, for each $\bm{\theta}^{(k)}$, independent MCMC chains are run targeting the posterior of the second module, $\pi(\bm{\alpha} \, | \, \bm{y}, \bm{\theta}^{(k)})$. The resulting pooled samples provide a Monte Carlo approximation to \eqref{eq:cut_function}. Furthermore, using Rubin's combining rules for multiple imputation \citep{Rubin1987} (see the Supplementary Material for details), we can quantify the total variance of an estimator as the sum of the outcome model variance (i.e., the ``within variance'') and the interference model variance (i.e., the ``between'' variance). This between/within assessment is an important advantage of this modular approach, as it allows us to quantify how much uncertainty in the causal effect estimates can be attributed to uncertainty in the interference structure. When uncertainty in the estimated interference structure dominates the uncertainty in the causal effect estimates, the interpretations of the effect estimates should be discussed in the context of the estimated interference structure.

In a simulation study designed to investigate the bias and coverage properties of the proposed estimator, we compare the above with a simpler ``plug-in'' estimator, in which the outcome model is fitted conditional on the posterior mean estimate of the process parameters, $\bar{\bm{\theta}} = E\widehat{(\bm{\theta} | \bar{\bm{\eta}})}$, i.e., uncertainty from the interference structure is \textit{not} propagated to the causal effect estimates. A detailed discussion of the simulation design and results is included in Supplementary Material. In general, we found that the two estimators exhibit similar levels of bias, however their coverage properties often differ. In particular, the plug-in estimator's posterior credible interval coverage rates were almost always lower than the estimator which incorporated uncertainty in the interference structure; this was especially pronounced when the outcome model was otherwise correctly specified. We also compared estimation using a log-linear BART response surface model (Section~\ref{sec:effects_est}) with a simple parametric alternative. Log-linear BART proved adept at estimating increasingly nonlinear dose-response surfaces, albeit with more conservative uncertainty quantification. Finally, Rubin's combining rules were used to quantify the variance in the estimates attributed to uncertainty in the interference structure; large differences in coverage rates between the two estimators were associated with a high proportion of variance attributed to the interference model.

\section{Effects of Scrubbers on Pediatric Asthma and Medicare Mortality in Texas}
\label{sec:results}

\begin{figure}[h]
\centering
     \begin{subfigure}[b]{0.3\textwidth}
         \centering
         \includegraphics[width=\textwidth]{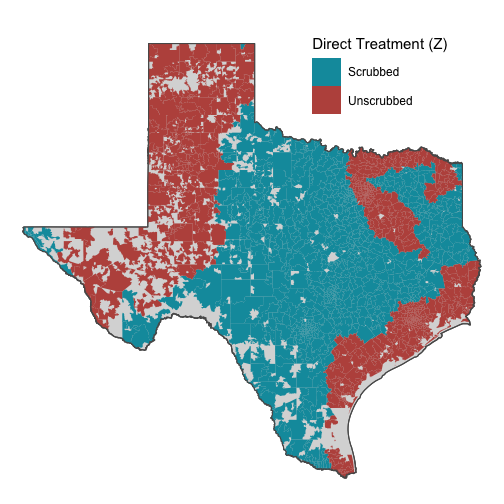}
         \caption{Key-associated treatment, $Z_i$.}
         \label{fig:z-dist}
     \end{subfigure}
     \hfill
     \begin{subfigure}[b]{0.3\textwidth}
         \centering
         \includegraphics[width=\textwidth]{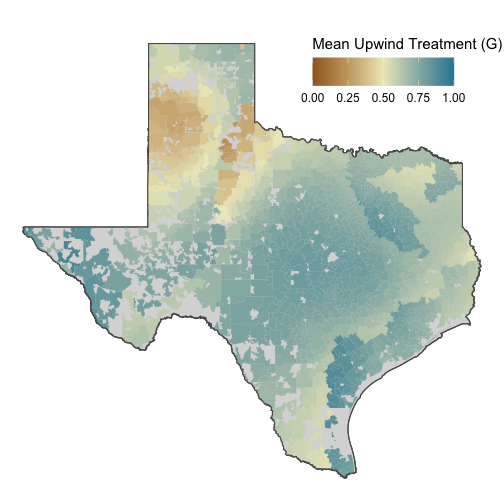}
         \caption{Mean upwind treatment, $\bar{G}_i$}
         \label{fig:g-mean}
     \end{subfigure}
     \hfill
     \begin{subfigure}[b]{0.3\textwidth}
         \centering
         \includegraphics[width=\textwidth]{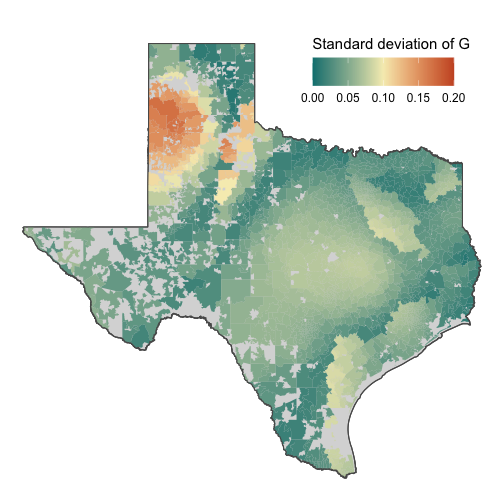}
         \caption{Standard deviations of $G_i$.}
         \label{fig:g-sd}
     \end{subfigure}
\caption{The key-associated and upwind treatment levels used in the analysis. Note that the distribution of $G$ is summarized using (b) the posterior mean estimate, $\bar{G}_i$, and (c) the marginal posterior standard deviation.}
\label{fig:treatments}
\end{figure}

Using the methods described in the preceding sections, we estimate the effect of scrubber presence at coal-fired power plants in 2016 on all-cause mortality among Medicare beneficiaries and pediatric asthma emergency department visits in Texas during that same year.  As described in Section~\ref{sec:sulfate_model}, the interference structure is estimated from a mechanistic model connecting power plant \ce{SO2} emissions to annual sulfate concentrations; the model is fitted to the observed 2016 average annual sulfate concentrations, \ce{SO2} emissions totals, and average (10m) wind velocities (Figure~\ref{fig:pollution_transport}). Given the estimated interference structure and the scrubber status of the 81 coal-fired power plants operating in the study region in 2016 (Figure~\ref{fig:scrubbers}), we assign key-associated and upwind treatment levels to 1935 Texas ZCTAs. Figure~\ref{fig:treatments} shows the spatial distribution of key-associated and upwind treatments across Texas: ~55\% of the ZCTAs are key-associated with a scrubbed facility, while the median value of $\bar{G}_i$ is 0.74 (range: 0.23--0.91).

Twenty eight covariates were used in the analysis (Table~\ref{tab:params}), including ZCTA-level demographic data obtained from the US census, climate data (such as annual total precipitation, average minimum and maximum daily temperature, and average relative humidity), annual mean black carbon concentrations, smoking prevalence, and power plant characteristics, including the annual operating time and heat input of the key-associated facility, distance to key-associated facility, and two summary statistics of the upwind facilities: the weighted degree, $T^{*}_{i\cdot} = \sum_{j \neq j^{*}_{(i)}} T_{ij}$, which quantifies ZCTA $i$'s potential exposure to \ce{SO4^{2-}} due to emissions from upwind power plants, and $\sum_{j \neq j^{*}_{(i)}} \text{heat}_j \, T_{ij} / T^{*}_{i\cdot}$, the weighted average of the upwind power plant heat inputs. 

The ZCTA demographic and climate features were selected for their known relevance to variation in pollution-related health endpoints, and the black carbon concentration as a general marker of urban pollution that is not related to power plants. Power plant characteristics (and their summary functions) are chosen to account for the possibility that certain types of plants are more or less likely to install scrubbers.  The inclusion of both of these types of covariates accounts for the possibility that power plants choose to install scrubbers based in part on the knowledge of downwind population features.  Further discussion of the types of confounding that should be considered in bipartite networks and with power plants specifically appears in \cite{Zigler2020}.

We implemented covariate balancing propensity scores (CBPS) \citep{Imai2014, Fong2018} for the sole purpose of evaluating covariate balance and overlap across treatment levels of $Z_i$ and $\bar{G}_i$. For binary treatment $Z_i$, we compared the covariates' absolute standardized mean differences between treated and untreated units \citep{Austin2009}, while balance for continuous treatment $\bar{G}_i$ was assessed using the Pearson correlation between covariate and treatment \citep{Austin2019}.  The results are displayed in the Supplementary Material: in general, unadjusted comparisons exhibited moderate to severe imbalance, however, these imbalances were largely resolved after propensity score adjustment. Similarly, propensity score overlap was achieved across treatment levels using CBPS. Even though inference for causal effects will not be based on the CBPS, this analysis indicates plausibility of the BART approach to adjust for observed confounding without extrapolation due to lack of overlap.

\begin{table}[ht]
\caption {Covariates included in the analysis, including demographic, climate, and facility-level data.} \label{tab:params}
\setlength{\tabcolsep}{12pt}
\renewcommand{\arraystretch}{1} 
\small 
\centering
\scalebox{0.9}{
	\begin{tabular}{lll}
	  \toprule
	  \textbf{Texas ZCTA Covariates} &  & \\ 
	  \midrule
    	  \textbf{\textit{US Census Data}} & 11. movement rate & 21. black carbon\\
	  1. total population  & 12. \% insured  & \textbf{\textit{Smoking Data}}  \\
    	  2. \% female & 13. \% renter housing & 22. Smoking rate \\
    	  3. \% kids ($0-17$ y.o.)  & 14. \% urban &  \textbf{\textit{Power Plant Data}} \\
    	  4. median age  & 15. population density & 23. key-assoc. operating time \\
    	  5. \% white  & \textbf{\textit{Climate Data}} & 24. heat input  \\
    	  6. \% black & 16. precipitation & 25. \% capacity \\
    	  7. \% hispanic & 17. minimum daily temp. & 26. distance to key-assoc. \\
    	  8. \% high school graduate & 18. maximum daily temp. & 27. upwind heat input \\
    	  9. \% below poverty level & 19. vapor pressure  & 28. weighted degree \\
    	  10. median income & 20. relative humidity  &  \\    
	  \bottomrule
	\end{tabular}
}
\end{table}

We considered two Poisson regressions to model $E(Y_i | \bm{X}_i, Z_i, G_i)$: (i) a log-linear BART model \eqref{eq:log_BART}, where the rate function, $f_{\bm{\alpha}}(\bm{x}_i, z_i, g_i)$, is defined as the log-linear sum of $m$ independent regression trees, and (ii) a parametric Poisson regression model with log-linear rate function, 
\begin{equation}
\log f_{LM}(\bm{x}_i, z_i, g_i) = \bm{x}_i' \bm{\beta} + \phi z_i + \gamma g_i + \psi z_i g_i.
\end{equation}
In both models, $\mu_{0i}$ is included as a population offset. Given the spatial structure of the outcome data, we used Moran's $I$ to test for spatial autocorrelation in the fitted models' residuals \citep{CliffOrd1981}. The tests found no significant evidence of spatial autocorrelation with either pediatric asthma ED visits or Medicare all-cause mortality as the outcome; this was true for both the parametric and nonparametric regression, and was invariant to the choice of spatial weights matrix or type of residual (see the Supplementary Material for details). Other settings that do exhibit spatial autocorrelation might benefit from the inclusion of a spatial random effect in the Poisson GLM or nontrivial extensions to the log-linear BART model, possibly following the example of \cite{Muller2007}.

Modular Bayesian inference was performed using both the interference uncertainty method described in Section~\ref{sec:prop_unc} and a simpler plug-in posterior mean estimate of the interference structure; the interference uncertainty method was fitted in parallel using 250 independent draws from $\pi(\bm{\theta} | \bar{\bm{\eta}})$. The log-linear BART models were assigned $m = 200$ additive trees, using the tree splitting rules and leaf prior tuning parameter specification recommended by \cite{Murray2021}. Posterior samples for all models were obtained using Metropolis-Hastings MCMC, and all analysis was conducted in R, version 4.1.3 (repository: \url{https://github.com/nbwikle/estimating-interference}).

\subsection{Medicare all-cause mortality}

\begin{figure}[ht]
    \centering
    \includegraphics[width=1\textwidth]{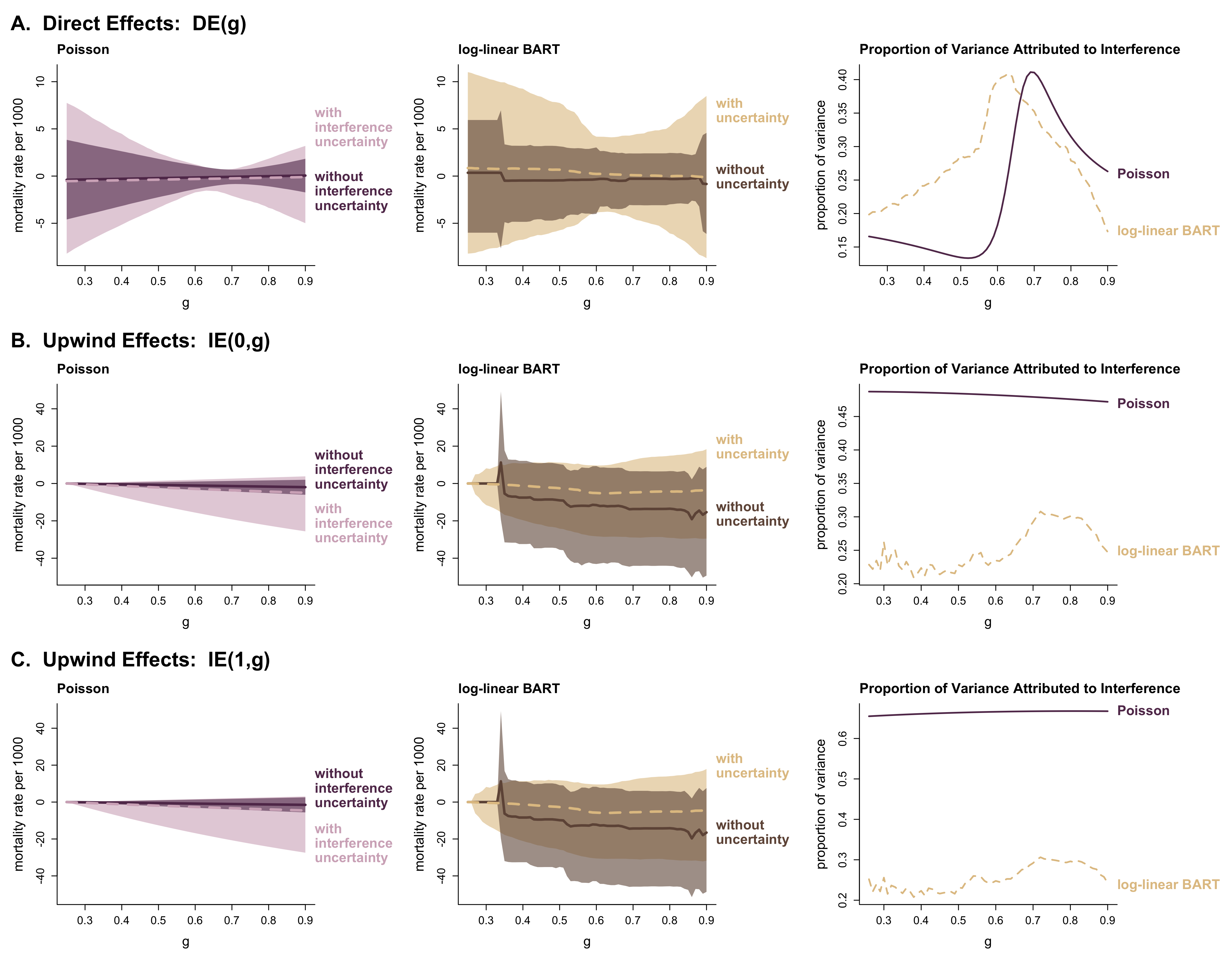}
    \caption{Estimated direct ($DE(g)$) and indirect ($IE(z,g)$) effects of coal-fired power plant scrubbers on the 2016 rate of all-cause mortality among Medicare beneficiaries in Texas.}
    \label{fig:medicare-results}
\end{figure}

The estimated direct and indirect (i.e., upwind) effects of scrubbers on all-cause mortality among Medicare beneficiaries are shown in Figure~\ref{fig:medicare-results}. Notably, there is little evidence that the presence of scrubbers on key-associated power plants had a significant effect on the rate of all-cause mortality among Texas Medicare beneficiaries in 2016 (Figure~\ref{fig:medicare-results}A). The estimates are similar when using either the parametric or log-linear BART outcome models, as well as either the plug-in or interference uncertainty methods. In contrast, the estimated indirect effects, $IE(z,g)$, vary based on the choice of outcome model and modular inference (Figures~\ref{fig:medicare-results}B and C). If we restrict our focus to the log-linear BART estimate with plug-in inference, we see some evidence that an increase in the proportion of scrubbed upwind power plants caused a reduction in all-cause mortality per 1000 Medicare beneficiaries (albeit with large uncertainty bounds). For example, in the absence of a scrubber at the closest power plant, the estimated effect of an increase in the proportion of scrubbed power plants from $g^{*} = 0.25$ to $\bar{g}_{0.5} = 0.74$ (the median average exposure level across all ZCTAs) is $\widehat{IE}(0, \bar{g}_{0.5}) = -13.6$ (95\% CI: $(-44.0, 6.5)$). However, the corresponding estimate accommodating uncertainty in the interference is $\widehat{IE}(0, \bar{g}_{0.5}) = -4.5$ $(-28.8, 13.9)$, and in general, the estimate of $\widehat{IE}(z,g)$ with interference uncertainty has shifted towards zero. In other words, the decision to propagate uncertainty in the interference structure had a substantial impact on the magnitude and interpretation of the estimated causal effects. We see similar results when comparing the parametric Poisson estimates --- there is little evidence that an increased proportion of upwind scrubbers had a significant effect on the rate of all-cause mortality among Texas ZCTAs in 2016. However, note that the uncertainty bounds for the estimates that propagate uncertainty in the interference structure are much wider than the plug-in alternatives, particularly in the parametric Poisson case when the proportion of variance attributable to the interference structure is large compared to that from the BART models.

\subsection{Pediatric asthma ED visits}

\begin{figure}[ht]
    \centering
    \includegraphics[width=1\textwidth]{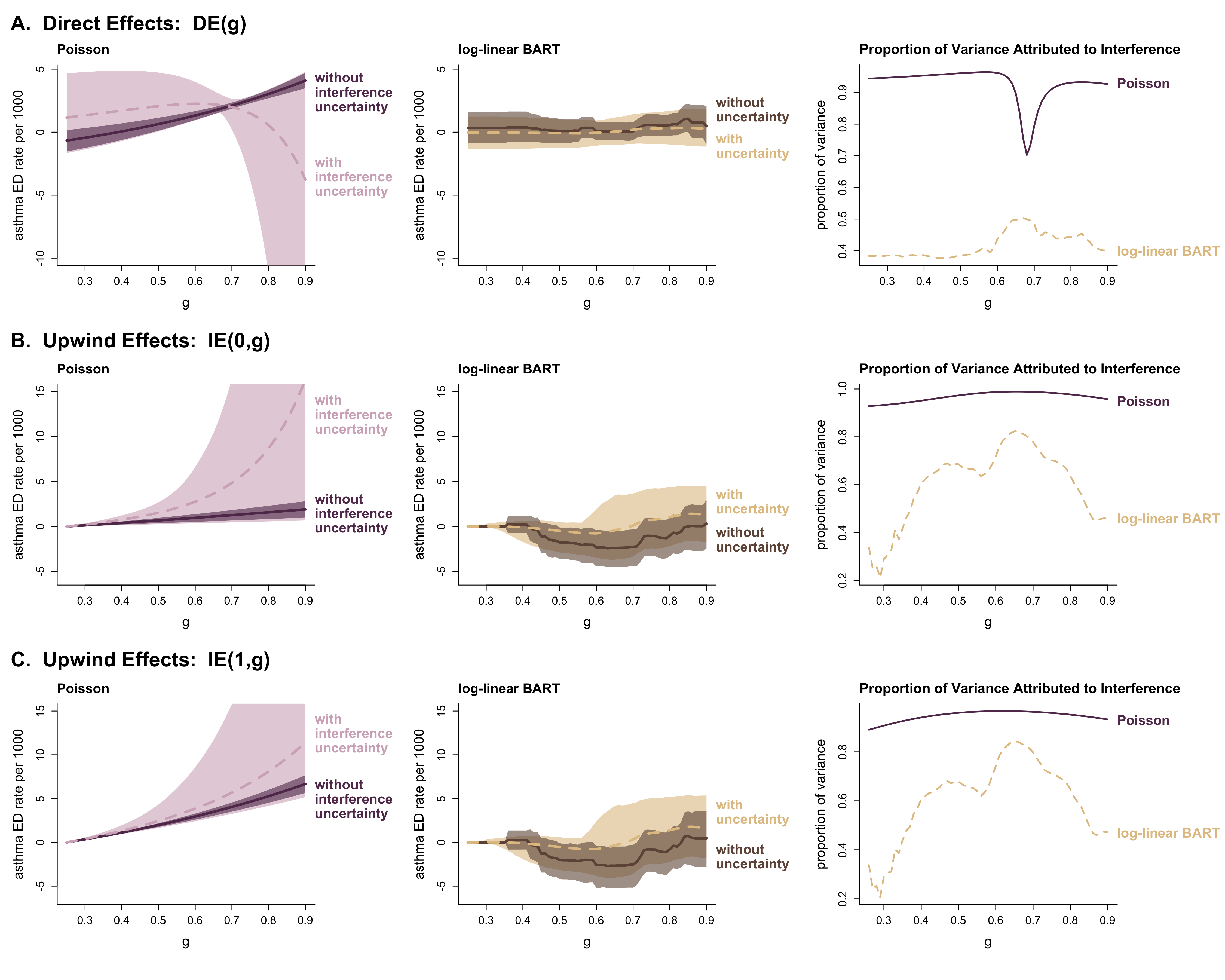}
    \caption{Estimated direct ($DE(g)$) and indirect ($IE(z,g)$) effects of coal-fired power plant scrubbers on the 2016 rate of pediatric asthma ED visits in Texas.}
    \label{fig:asthma-results}
\end{figure}

We performed a similar analysis of the impact of coal-fired power plant scrubbers on the rate of pediatric asthma-related ED visits in Texas in 2016; the results are shown in Figure~\ref{fig:asthma-results}. First, consider the log-linear BART plug-in estimates: $\widehat{DE}(g)$ is concentrated around zero, providing little evidence that the presence of a scrubber at the nearest power plant facility had a significant effect on the rate of pediatric asthma ED visits in 2016.  In contrast, the estimated indirect effect curves suggest that the rate of pediatric asthma ED visits was reduced as the proportion of scrubbed upwind power plants increased from 0.25 to 0.7, for example, $\widehat{IE}(0, 0.7) = -2.8$ (95\% CI: $(-5.6, -0.4)$). However, once again the addition of uncertainty from the interference estimates advises caution --- after accounting for uncertainty in the interference structure, it is no longer clear that an increase in the proportion of scrubbed upwind power plants caused a noticeable reduction in the 2016 rate of pediatric ED visits in Texas. Finally, we note that the parametric Poisson estimates differ substantially from the BART estimates, and in some cases, they suggest that the presence of scrubbers led to an increase in the rate of ED visits. However, there is reason for skepticism of the parametric estimates: the high proportion of variance attributed to uncertainty in the interference structure, combined with the very large uncertainty bounds around the interference uncertainty method's estimates, suggests that the Poisson regression model was not flexible enough to correctly characterize the response surface. In contrast, the log-linear BART model accommodates nonlinearity and interactions between confounders and treatment variables, contributing to the difference in point estimates and uncertainty bounds when compared with the parametric alternative.

\section{Discussion}
\label{sec:disc}

To our knowledge, this analysis represents the first example of an observational study in which the physical mechanism for interference has been estimated from available ancillary data. Our methods are especially relevant when evaluating the effectiveness of point-source air pollution interventions on downwind health outcomes: the impact of a change in emissions is likely non-local, as its effect on downwind pollution concentrations is dependent on the advection and reaction of pollutants across a wide spatio-temporal domain. Consequently, understanding of the pollution transport dynamics is necessary if we wish to characterize the spillover effects of multiple interventions across space. We showed how the mechanistic statistical model developed by \cite{Wikle2022} can be used to estimate the process dynamics from annual average sulfate concentrations; the estimated dynamics defined a weighted bipartite network linking interventional units to outcome locations, and a corresponding (probabilistic) exposure model was defined on the bipartite network. The need to estimate the dynamics is accompanied by inherent uncertainty, which we accommodate by averaging causal estimates over the range of possible interference structures as indicated by the posterior distribution of the spatio-temporal process. 

As with all observational studies, the results from our analysis should be considered in the context of the broader literature on the health impacts of air pollution exposure. For example, a number of studies have linked short and long-term pollution exposure with reduced pediatric lung function \citep{Garcia2021}. while exposure to \ce{SO4^{2-}} and \ce{PM_{2.5}} is associated with a variety of adverse health outcomes \citep{Dominici2014}, including increased risk of mortality \citep{Pope2009}, although the literature on health impacts attributable to \ce{PM_{2.5}} derived specifically from coal-fired power plants is evolving \citep{Henneman2023, Henneman2019b}. Consequently, it is worth asking why the estimated direct and indirect effects in Section~\ref{sec:results} are not more pronounced. First, we caution that the estimated causal effects are limited to the spatial and temporal extent of our study (i.e., observations from Texas ZCTAs in 2016). Furthermore, we note that estimating the health impacts of \textit{specific interventions} intended to control point-source pollution involves considerable difficulties and uncertainties --- such as understanding the spatio-temporal dynamics and complex interactions of the pollution transport process --- that are often not present in the more common task of estimating health impacts of (locally-measured) ambient pollution itself.

Our estimates are most comparable to those of \cite{Zigler2020}, who estimate the causal effects of emissions controls on Medicare ischemic heart disease (IHD) hospitalizations in the eastern US during 2005, where coal power plant pollution is more prominent than in Texas during 2016 \citep{Henneman2019}. As with our analysis, \cite{Zigler2020} did not find significant evidence of a direct effect (i.e., the effect of key-associated scrubbers on IHD hospitalizations). However, they did identify a significant reduction in IHD hospitalizations caused by an increase in upwind treatments. In contrast, the estimates from our analysis had wider uncertainty bounds, especially when including uncertainty in the estimated interference structure. We hypothesize that this may be due to (i) health outcomes (i.e., pediatric asthma and all-cause mortality) that are less affected by scrubbers than IHD hospitalizations, (ii) the smaller spatial and temporal extent used in our analysis compared to similar epidemiological studies \citep[e.g.,][]{Henneman2023} (iii) a time frame of study that took place when power plant pollution exposures were reduced relative to earlier years, or (iv) the incorporation of uncertainty in the interference structure. Finally, we note that although our analysis has adjusted for many potential confounders, the regulatory and market incentives driving some power plants to install scrubbers and the process dictating how these decisions propagate into pollution exposures remains incompletely understood, presenting, as with all observational studies, the lingering threat of unmeasured confounding.

This paper focuses on the estimation of an interference network originating from a complex physical system, and its success depends on a number of components, including the proposed bipartite causal inference framework, a mechanistic statistical model of air pollution transport, modularized Bayesian inference, modern extensions to Bayesian nonparametric modeling of count data, and the satisfaction of causal assumptions across different values of the exposure mapping dictated by uncertainty in the physical process model. Consequently, there are several possible extensions to this work. Chief among them is its extension to the spatio-temporal setting --- the installation of scrubbers changes over time, as do air pollution dynamics. Thus, the estimation of a spatio-temporal model of air pollution transport, and the corresponding time-varying interference structure, would lead to a more complete understanding of the effects of air pollution controls on health outcomes. In addition, extensions of this work to settings with higher-resolution outcome data, such as point-referenced outcomes, may allow for inferences that are more robust to the problems of ecological fallacy common when using spatially aggregated health data \citep{Diggle1995}. Other natural extensions might include adapting log-linear BART to target regularization-induced confounding \citep{Hahn2020} and better accommodate the estimation of heterogeneous effects, pursuing computational alternatives to the modularized Bayesian inference, or considering other strategies for confounding adjustment such as those based on (generalized) propensity scores. Furthermore, our choice of direct and indirect treatment, $Z_i$ and $G_i$, corresponds to only one of many possible exposure mappings, and alternative specifications relevant to other scientific questions of interest could be further explored (see, e.g., Section 7 of the Supplementary Material).

Above all, the work offered here is designed as an installment of methodology for observational studies with process-driven interference, which are of anticipated relevance for a variety of problems in the environmental and physical sciences. Dynamic spatio-temporal models \citep{WikleHooten2010} and scientific machine learning methods \citep{Rackauckas2021} are increasingly able to learn complex physical processes from massive scientific data sets. When possible, these estimated dynamics can help inform the topology linking interventions located at points or regions in space to relevant outcome units. Examples include assessing the impact of agricultural runoff on the health of downstream ecological systems \citep{Xia2020}, estimating the effect of hydraulic fracturing on the frequency and severity of earthquakes in the surrounding shale play \citep{McClure2017}, or attributing extreme weather events to climate change \citep{Wehner2023}. Furthermore, estimated dynamical processes may be of use in causal settings beyond the bipartite interference framework considered in this paper, perhaps suggesting spatially heterogeneous extensions to the spillover estimands of \cite{Wang2023} and \cite{Pollmann2023}, as well as a framework for causal inference with process-driven stochastic interventions \citep{DiazMunoz2012}. Ultimately, we believe that the continued development of process-informed causal methods is an important component in the creation and evaluation of effective health and environmental policies.

\section*{Acknowledgements}

This work was supported by research funding from NIH R01ES026217, R01ES034803, and US EPA 83587201. Its contents are solely the responsibility of the grantee and do not necessarily represent the official views of the USEPA. Further, USEPA does not endorse the purchase of any commercial products or services mentioned in the publication. 

We are grateful to Jared Murray for his helpful discussion and code base for the log-linear BART implementation. We also thank three anonymous reviewers for their helpful comments. The authors acknowledge the Texas Advanced Computing Center (TACC) at The University of Texas at Austin for providing HPC resources that have contributed to the research results reported within this paper. URL: \url{http://www.tacc.utexas.edu}

\section*{Supplementary Material}

Supplementary materials can be found at \url{https://github.com/nbwikle/estimating-interference}. This includes a discussion of the discrete space approximation of the sulfate model, an assessment of covariate balance and overlap, a detailed simulation study of the proposed estimator, a sensitivity study of the log-linear BART tuning parameters, replication of the analysis without a key-associated treatment, and a test for spatial autocorrelation. We have also included all code and any publicly available data from the analysis.

\bibliography{WikleZigler-references.bib}

\end{document}